\documentclass[aps,prl,twocolumn,superscriptaddress,showpacs,nobalancelastpage]{revtex4}
\usepackage{graphicx}
\begin{document}
\title{ Beam polarization asymmetries for the 
$p(\gamma,K^{+})\Lambda$ and $p(\gamma,K^{+})\Sigma^{0}$ reactions at $E_{\gamma}=1.5-2.4$ GeV}

\author{R.G.T.~Zegers}
  \affiliation{Research Center for Nuclear Physics, Osaka University, Ibaraki, Osaka 567-0047, Japan}
\author{M.~Sumihama}
  \affiliation{Department of Physics, Osaka University, Toyonaka, Osaka 560-0043, Japan}
  \affiliation{Advanced Science Research Center, Japan Atomic Energy Research Institute, Tokai, Ibaraki 319-1195, Japan}
\author{D.S.~Ahn}
  \affiliation{Department of Physics, Pusan National University, Busan 609-735, Korea}
\author{J.K.~Ahn}
  \affiliation{Department of Physics, Pusan National University, Busan 609-735, Korea}
\author{H.~Akimune}
  \affiliation{Department of Physics, Konan University, Kobe, Hyogo 658-8501, Japan}
\author{Y.~Asano}
  \affiliation{Synchrotron Radiation Research Center, Japan Atomic Energy Research Institute, Mikazuki, Hyogo 679-5198, Japan}
  \affiliation{Advanced Science Research Center, Japan Atomic Energy Research Institute, Tokai, Ibaraki 319-1195, Japan}
\author{W.C.~Chang}
  \affiliation{Institute of Physics, Academia Sinica, Taipei 11529, Taiwan}
\author{S.~Dat\'e}
  \affiliation{Japan Synchrotron Radiation Research Institute, Mikazuki, Hyogo 679-5198, Japan}
\author{H.~Ejiri}
  \affiliation{Japan Synchrotron Radiation Research Institute, Mikazuki, Hyogo 679-5198, Japan}
  \affiliation{Research Center for Nuclear Physics, Osaka University, Ibaraki, Osaka 567-0047, Japan}
\author{H.~Fujimura}
  \affiliation{School of Physics, Seoul National University, Seoul, 151-747, Korea}
\author{M.~Fujiwara}
  \affiliation{Research Center for Nuclear Physics, Osaka University, Ibaraki, Osaka 567-0047, Japan}
  \affiliation{Advanced Science Research Center, Japan Atomic Energy Research Institute, Tokai, Ibaraki 319-1195, Japan}
\author{K.~Hicks}
  \affiliation{Department of Physics and Astronomy, Ohio University, Athens, Ohio 45701}
\author{T.~Hotta}
  \affiliation{Research Center for Nuclear Physics, Osaka University, Ibaraki, Osaka 567-0047, Japan}
\author{K.~Imai}
  \affiliation{Department of Physics, Kyoto University, Kyoto 606-8502, Japan} 
\author{T.~Ishikawa}
  \affiliation{Laboratory of Nuclear Science, Tohoku University, Sendai, Miyagi 982-0826, Japan}
\author{T.~Iwata}
  \affiliation{Department of Physics, Yamagata University, Yamagata 990-8560, Japan}
\author{H.~Kawai}
  \affiliation{Department of Physics, Chiba University, Chiba 263-8522, Japan}
\author{Z.Y.~Kim}
  \affiliation{School of Physics, Seoul National University, Seoul, 151-747, Korea}
\author{K.~Kino}
  \affiliation{Research Center for Nuclear Physics, Osaka University, Ibaraki, Osaka 567-0047, Japan}
\author{H.~Kohri}
  \affiliation{Research Center for Nuclear Physics, Osaka University, Ibaraki, Osaka 567-0047, Japan}
\author{N.~Kumagai}
  \affiliation{Japan Synchrotron Radiation Research Institute, Mikazuki, Hyogo 679-5198, Japan}
\author{S.~Makino}
  \affiliation{Wakayama Medical University, Wakayama, Wakayama 641-8509, Japan}
\author{T.~Matsumura}
  \affiliation{Research Center for Nuclear Physics, Osaka University, Ibaraki, Osaka 567-0047, Japan}
  \affiliation{Advanced Science Research Center, Japan Atomic Energy Research Institute, Tokai, Ibaraki 319-1195, Japan}
\author{N.~Matsuoka}
  \affiliation{Research Center for Nuclear Physics, Osaka University, Ibaraki, Osaka 567-0047, Japan}
\author{T.~Mibe}
  \affiliation{Research Center for Nuclear Physics, Osaka University, Ibaraki, Osaka 567-0047, Japan}
  \affiliation{Advanced Science Research Center, Japan Atomic Energy Research Institute, Tokai, Ibaraki 319-1195, Japan}
\author{K.~Miwa}
  \affiliation{Department of Physics, Kyoto University, Kyoto 606-8502, Japan} 
\author{M.~Miyabe}
  \affiliation{Department of Physics, Kyoto University, Kyoto 606-8502, Japan} 
\author{Y.~Miyachi}
  \affiliation{Department of Physics and Astrophysics, Nagoya University, Nagoya, Aichi 464-8602, Japan}
\author{M.~Morita}
  \affiliation{Research Center for Nuclear Physics, Osaka University, Ibaraki, Osaka 567-0047, Japan}
\author{N.~Muramatsu}
  \affiliation{Advanced Science Research Center, Japan Atomic Energy Research Institute, Tokai, Ibaraki 319-1195, Japan}
\author{T.~Nakano}
  \affiliation{Research Center for Nuclear Physics, Osaka University, Ibaraki, Osaka 567-0047, Japan}
\author{M.~Niiyama}
  \affiliation{Department of Physics, Kyoto University, Kyoto 606-8502, Japan} 
\author{M.~Nomachi}
  \affiliation{Department of Physics, Osaka University, Toyonaka, Osaka 560-0043, Japan}
\author{Y.~Ohashi}
  \affiliation{Japan Synchrotron Radiation Research Institute, Mikazuki, Hyogo 679-5198, Japan}
\author{T.~Ooba}
  \affiliation{Department of Physics, Chiba University, Chiba 263-8522, Japan}
\author{H.~Ohkuma}
  \affiliation{Japan Synchrotron Radiation Research Institute, Mikazuki, Hyogo 679-5198, Japan}
\author{D.S.~Oshuev}
  \affiliation{Institute of Physics, Academia Sinica, Taipei 11529, Taiwan}
\author{C.~Rangacharyulu}
  \affiliation{Department of Physics and Engineering Physics, University of Saskatchewan, Saskatoon, Saskatchewan, Canada, S7N 5E2} 
\author{A.~Sakaguchi}
  \affiliation{Department of Physics, Osaka University, Toyonaka, Osaka 560-0043, Japan}
\author{T.~Sasaki}
  \affiliation{Department of Physics, Kyoto University, Kyoto 606-8502, Japan} 
\author{P.M.~Shagin}
  \affiliation{Research Center for Nuclear Physics, Osaka University, Ibaraki, Osaka 567-0047, Japan}
\author{Y.~Shiino}
  \affiliation{Department of Physics, Chiba University, Chiba 263-8522, Japan}
\author{H.~Shimizu}
  \affiliation{Laboratory of Nuclear Science, Tohoku University, Sendai, Miyagi 982-0826, Japan}
\author{Y.~Sugaya}
  \affiliation{Department of Physics, Osaka University, Toyonaka, Osaka 560-0043, Japan}
\author{H.~Toyokawa}
  \affiliation{Japan Synchrotron Radiation Research Institute, Mikazuki, Hyogo 679-5198, Japan}
\author{A.~Wakai}
  \affiliation{Center for Integrated Research in Science and Engineering, Nagoya University, Nagoya, Aichi 464-8603, Japan}
\author{C.W.~Wang}
  \affiliation{Institute of Physics, Academia Sinica, Taipei 11529, Taiwan}
\author{S.C.~Wang}
  \affiliation{Institute of Physics, Academia Sinica, Taipei 11529, Taiwan}
\author{K.~Yonehara}
  \affiliation{Department of Physics, Konan University, Kobe, Hyogo 658-8501, Japan}
\author{T.~Yorita}
  \affiliation{Japan Synchrotron Radiation Research Institute, Mikazuki, Hyogo 679-5198, Japan}
\author{M.~Yoshimura}
  \affiliation{Institute for Protein Research, Osaka University, Suita, Osaka 565-0871, Japan}
\author{M.~Yosoi}
  \affiliation{Department of Physics, Kyoto University, Kyoto 606-8502, Japan} 
\collaboration{The LEPS collaboration}
  \noaffiliation

\date{\today}

\begin{abstract}
Beam polarization asymmetries for the $p(\vec{\gamma},K^{+})\Lambda$ and $p(\vec{\gamma},K^{+})\Sigma^{0}$ reactions are measured for the first time for $E_{\gamma}=1.5-2.4$ GeV and $0.6<\cos(\theta_{K^{+}}^{cm})<1.0$ by using linearly polarized photons at the Laser-Electron-Photon facility at SPring-8 (LEPS). The observed asymmetries are positive and gradually increase with rising photon energy. The data are not consistent with theoretical predictions based on tree-level effective Lagrangian approaches. Including the new results in the development of the models is, therefore, crucial for understanding the reaction mechanism and to test the presence of baryon resonances which are predicted in quark models but are sofar undiscovered.
\end{abstract}


\maketitle

Strangeness photoproduction is a powerful tool to obtain a deeper
insight into baryon resonances. 
It provides additional information about the baryon resonances to that
obtained from $\pi N$ scattering and $\pi$-production reactions. 
Of special interest are nucleon resonances that have been predicted in quark
models \cite{CAP86} and for which no experimental evidence has been
found via the $\pi$-induced or $\pi$-production reactions. Some of
these resonances, referred to as `missing', could couple strongly to
the $K\Lambda$ and $K\Sigma$ channels \cite{CAP94,CAP98}. To better
understand the problem of `missing' resonances and to see whether predictions of baryon resonances can be tested, it is,
therefore, very interesting to study experimentally the $p(\gamma,K^{+})\Lambda$ and $p(\gamma,K^{+})\Sigma^{0}$ reactions.

Measurements of the energy dependence of the total cross section for the $p(\gamma,K^{+})\Lambda$ reaction at SAPHIR/Bonn \cite{TRA98} resulted in renewed interest because of the presence of a resonance-like structure near $W=1900$ MeV. Mart and Bennhold showed that this structure could be explained by introducing a $D_{13}(1895)$ resonance \cite{MAR00} for which a considerable branching into the $K\Lambda$ channel is predicted \cite{CAP98}. Measurements of the cross section at CLAS/JLAB \cite{SCH02} suggest that the resonance-like structure actually consists of several components which manifest themselves at different $K^{+}$-scattering angles. 

The theoretical calculations are typically performed in a tree-level effective-Lagrangian approach. Janssen \textit{et al.} showed, however, that large ambiguities arise from (i) the choice of included resonances, (ii) coupling constants, (iii) form factors and (iv) the treatment of the non-resonant `background' \cite{JAN01,JAN02}. Great caution is thus advised in drawing definite conclusions based on the cross-section data only. Alternative theoretical approaches in which, for example, off-shell effects are taken into account \cite{SAG01}, can also describe the SAPHIR data well without inclusion of `missing' resonances. Moreover, coupled-channels effects are not negligible \cite{CHI01}. One way to limit the freedoms in the model calculations is to analyze results from all photon-induced channels simultaneously \cite{PEN02}. 

For the development of the models it is of vital importance to measure additional observables and improve the quality of the cross section data. Results for the recoil-polarization asymmetry in the $p(\gamma,K^{+})\Lambda$ reaction (self analyzing by the $\Lambda$ weak decay) are already available from the SAPHIR data set. Extensive programs to measure cross sections and recoil polarizations are underway at JLAB/CLAS \cite{SCH02} and ESRF/GRAAL \cite{BOC01}. 
Additionally, measurements of the beam polarization asymmetry ($\Sigma$) are great assets to the database because of the high sensitivity to the model parameters and the presence of resonances \cite{MAR00,JAN01}. This asymmetry is defined through $(\frac{d\sigma}{d\Omega})_{pol}=\frac{d\sigma}{d\Omega}[1+P_{\gamma}\Sigma\cos(2\phi')]$,
where $(\frac{d\sigma}{d\Omega})_{pol}$ is the cross section using a linearly-polarized photon beam, $\frac{d\sigma}{d\Omega}$ is the unpolarized cross section, $P_{\gamma}$ the degree of photon polarization, $\phi'$ the azimuthal angle between the photon polarization plane and the vector normal to the $K^{+}$ reaction plane. Access to this observable is most easily obtained at backward-Compton scattering facilities \cite{BAR02,NAK01} because the photon beam is easily and reliably polarized to a high degree.    

\begin{figure}
\scalebox{1.0}{\includegraphics{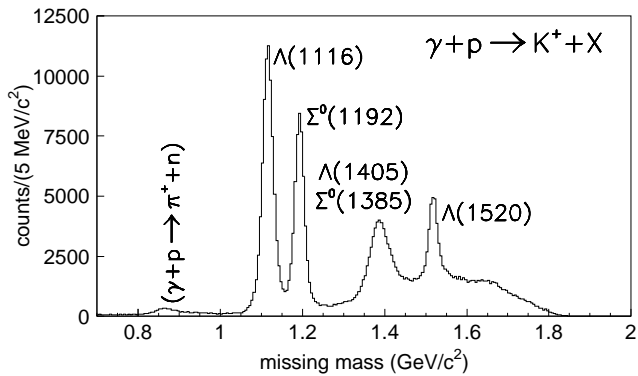}}
\caption{\label{fig:prl1} Missing-mass spectrum for the $p(\gamma,K^{+})X$ reaction.}
\end{figure}

In this Letter, we present for the first time measurements of the 
beam polarization asymmetries of the $p(\gamma,K^{+})\Lambda$ and
$p(\gamma,K^{+})\Sigma^{0}$ reactions. These data were taken at the new SPring8/LEPS facility in Japan \cite{NAK01}. Photons with
a maximum energy of 2.4 GeV were produced from backward Compton
scattering of 351-nm laser photons off 8-GeV electrons in the SPring-8
storage ring. The photons were tagged by measuring the scattered
electron energies with a resolution $\sigma$=15 MeV. The degree of
polarization of the backscattered photon beam was 95\% at 2.4 GeV and
55\% at 1.5 GeV. Half of the data was taken with
horizontally-polarized photons and the other half with
vertically-polarized photons. The direction of the polarization was
switched about every 2 hours.
The typical photon flux was $10^{6}$/s. A 50-mm thick liquid-hydrogen target was used. 

Charged particles were momentum-analyzed by tracing their paths in a
magnetic dipole field by means of a silicon-strip vertex detector and
one drift chamber positioned upstream from the dipole magnet, and two
drift chambers positioned downstream of the dipole magnet. The
upstream drift chamber consists of 6 wire planes (3 vertical planes, 2
planes at $+45^{\circ}$ and 1 plane at $-45^{\circ}$) and each of the
downstream drift chambers consists of 5 wire planes (2 vertical
planes, 2 planes at $+30^{\circ}$ and 1 plane at
$-30^{\circ}$). Electron and positron tracks due to pair production
were largely removed at the trigger level by means of an aerogel
\v{C}erenkov veto counter. The event sample was further cleaned up by
removing tracks with a large track-reconstruction error (confidence
level $<2$\%), which were mostly due to decay-in-flight events. The
time-of-flight of each track was measured; the start signal was
produced by a plastic-scintillator trigger counter placed behind the
target cell, and an array of 40 plastic scintillators placed behind
the tracking detectors provided the stop signal. The time-of-flight
resolution was about 150 ps for a typical path length of 4 m. By
combining time-of-flight and momentum, the mass of each track was
reconstructed with a resolution ($\sigma$) of 30 (105) MeV/$c^{2}$ for
a 1 (2) GeV/$c$ kaon. A $3\sigma$-mass cut was used to select the
positively-charged kaons, with the additional condition that
$0.31<\text{mass}<0.74$ GeV/c$^{2}$ to ensure that the $K^{+}$
cut does not overlap with the cuts for the $\pi^{+}$ and proton.  At
the highest momenta ($\sim 2$ GeV/c), where the mass resolution was
worst, the contamination from the $\pi^{+}$-particles and protons
amounted to 2\% (3.5\%) and 2.5\% (5\%) for the $K^{+}\Lambda$
($K^{+}\Sigma^{0}$) production, respectively. These numbers were
determined by extrapolating the Gaussian-shaped mass distributions of
the $\pi^{+}$'s and protons into the $K^{+}$ region. $K^{+}$-mesons scattered between $0^{\circ}$ and $60^{\circ}$ degrees in the center-of-mass frame were detected by the LEPS detector \cite{NAK01}. The track-angle resolution was 2.3 mrad.
\begin{figure}
\scalebox{1.0}{\includegraphics{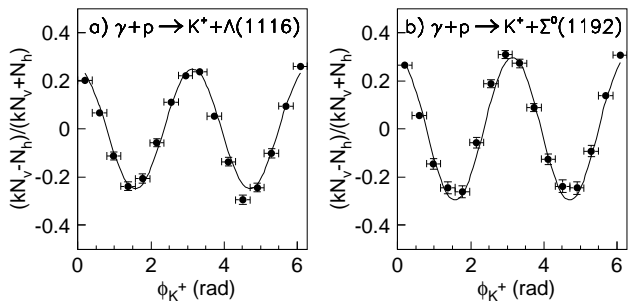}}
\caption{\label{fig:asym} Asymmetry spectra for the $p(\gamma,K^{+})\Lambda$ (a) and $p(\gamma,K^{+})\Sigma^{0}$ (b) reactions for all events. A fit to the data with $C\cos{2\phi}$ is superimposed.}
\end{figure}

Fig. \ref{fig:prl1} shows the missing-mass spectrum obtained for the
$p(\gamma,K^{+})X$ reaction. Besides $\Lambda(1116)$ and
$\Sigma^{0}(1193)$, additional peaks due to $\Lambda(1405)$,
$\Sigma^{0}(1385)$ (the two are not resolved) and $\Lambda(1520)$ are
observed. A small bump below 1 GeV/c$^{2}$ is due to misidentified
$\pi^{+}$ tracks. The missing-mass resolutions for the $\Lambda$
($\Sigma^{0}$) were $\sigma =$17(16) and 10(9) MeV/c$^{2}$ at the
highest and lowest momenta, respectively. A momentum-dependent $2\sigma$ cut was used to select the events in each peak. The contamination of $\Lambda$ ($\Sigma^{0}$) events in the $\Sigma^{0}$ ($\Lambda$) peak is less than 0.8\% (0.4\%). In total, $7.3\times 10^4$ $K^{+}\Lambda$ and $4.9\times 10^4$ $K^{+}\Sigma^{0}$ events satisfied all conditions given above.

The beam polarization asymmetries ($\Sigma$) are determined using the relation:
\begin{equation}
\label{eq:ba}
\Sigma P_{\gamma} \cos(2\phi) = \frac{kN_{v}(\phi)-N_{h}(\phi)}{kN_{v}(\phi)+N_{h}(\phi)},
\end{equation}
where $N_{v}$ ($N_{h}$) is the number of events detected at angle
$\phi$, with a vertically (horizontally) polarized photon beam and $k$
is a normalization factor obtained from the integrated photon yield
for each polarization mode, corrected for the dead-time of the
data-acquisition system and the random tagger-hit rate. The azimuthal
angle $\phi$ is measured with respect to the horizontal plane. Note
that the detector acceptance is not present in Eq. \ref{eq:ba}, which
is valid because the acceptances for our data taken with a horizontally and
vertically-polarized photon beam are very nearly the same. Fig. \ref{fig:asym}
shows the measured ratio in the r.h.s. of Eq. \ref{eq:ba} for the
total $K^{+}\Lambda$ (a) and $K^{+}\Sigma^{0}$ (b) samples. By fitting
with a $C\cos(2\phi)$ function  and dividing $C$ by $P_{\gamma}$,
$\Sigma$ is obtained. When using the full data sets, the statistical
errors are smaller than the systematic ones (see below). 

The $K^{+}\Lambda$  and $K^{+}\Sigma^{0}$ data sets were each divided
into 9, 0.1-GeV wide, photon-energy bins ranging from 1.5 to 2.4 GeV. The narrow energy binning is important, since the excitation
spectrum may vary rapidly due to the presence of resonances. The
chosen bin-size is smaller than or comparable to the widths of the
relevant baryon resonances. For each energy bin, the events were
further divided according to $K^{+}$ scattering angles; 5 bins in
$\cos(\theta_{K^{+}}^{cm})$ from 0.6 to 1.0, each with a width of 0.1,
except for the 2 most forward bins which had a width of 0.05. For each
sub-sample, the beam polarization asymmetry was determined following
the above-described procedure (the reduced $\chi^{2}$ of the fits with a $\cos(2\phi)$ to the measured asymmetries varied from 0.4 to 2.1). Although the contamination from protons
and $\pi^{+}$'s in the $K^{+}$ sample was small, it gives rise to a
non-negligible shift of the measured asymmetry for the $K^{+}$. This
was corrected for by determining contamination level from the protons
and $\pi^{+}$'s and their respective asymmetries (determined by
selecting $\pi^{+}$'s and protons in the mass spectra but keeping all
other selections described above; for protons the asymmetries are close to
0 and for $\pi^{+}$'s they are positive, but in general slightly lower than for
the $K^{+}$'s). Since the asymmetry of the total sample is the average of the asymmetries for the $K^{+}$ events and the proton and $\pi^{+}$ contaminations, weighted by their relative contributions in each sample, the asymmetries for the $K^{+}$-sample can be extracted. The correction ranged from $0.00\pm 0.01$ for the lowest photon energies to $+0.03\pm 0.02$ at the highest photon energies. 

\begin{figure}
\scalebox{0.85}{\includegraphics{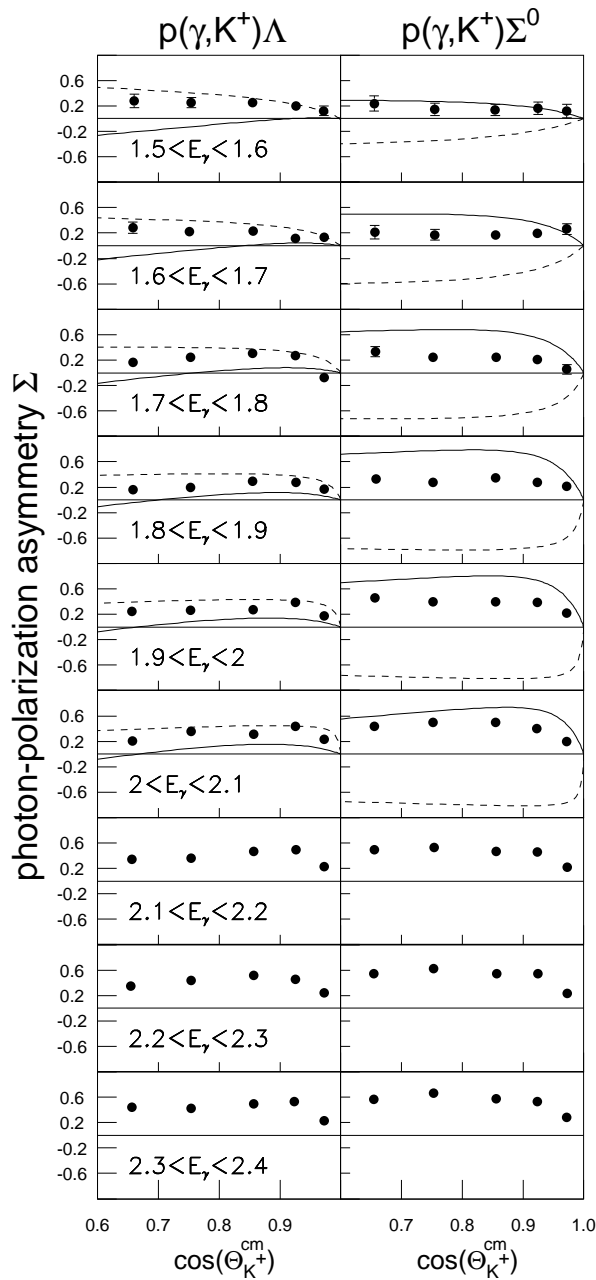}}
\caption{\label{fig:sigm}Beam polarization asymmetries for the $p(\gamma,K^{+})\Lambda$ (left) and $p(\gamma,K^{+})\Sigma^{0}$ (right) reactions as a function of $\cos(\theta_{K^{+}}^{cm})$ for different photon-energy bins. The error bars are mostly smaller than the markers. Theoretical predictions using the MAID2000 program \cite{KMA} (dashed lines) and by Janssen \textit{et al.} \cite{JAN01,JAN02} (solid lines) are compared with the experimental data.}
\end{figure}

The final results are shown in Fig. \ref{fig:sigm}. The observed
asymmetries are positive and increase gradually with rising photon
energy. The error bars correspond to the combined statistical (ranging
from 0.09 at $E_{\gamma}=1.5$ GeV to 0.04 at $E_{\gamma}=2.4$ GeV) and
systematic errors ($\sim 0.02$). The latter arise from (i) the
photon-yield normalization errors ($k$ in Eq. \ref{eq:ba}) and the
uncertainties in the degree and angle of linear polarization
(systematic error: 0.01), (ii) the partial loss of events in a subset of the data due to a trigger problem in case the decay proton from the $\Lambda$ ($\Lambda\rightarrow
p\pi^{-}$ or $\Sigma^{0}\rightarrow\Lambda\gamma$, $\Lambda
\rightarrow p\pi^{-}$) hit the trigger counter. The loss is slightly dependent on the polarization direction and the effect on the measured asymmetries was estimated by mimicking the trigger problem in the subset of the data where it did not occur (systematic error
(0.01 (0.015) for $\Lambda$ ($\Sigma^{0}$) production), (iii) contamination from events produced at the trigger counter, which is only significant at very forward  $K^{+}$ scattering angles ($\cos(\theta_{K^{+}}^{cm})>0.95$); the systematic error is negligible for $\Lambda$ production and 0.01 for $\Sigma^{0}$ production). 

In Fig. \ref{fig:sigm} the experimental data are compared with the theoretical predictions using the MAID2000 program \cite{MAR00,LEE00,KMA} (dashed lines) and by Janssen \textit{et al.} \cite{JAN01,JAN02} (solid lines). These calculations are the most up-to-date available and good examples to see model ambiguities and the sensitivity of the beam polarization asymmetry on the model assumptions. Both calculations are obtained on the basis of a tree-level effective Lagrangian model and make use of the cross-section data from SAPHIR to fix the various parameters in the models through a fitting procedure. The same $s$-channel resonances are taken into account, including the `missing' $D_{13}(1895)$ resonance. With the $D_{13}(1895)$ resonance, the calculations reproduce the experimental cross sections better but also give dramatically different predictions for the beam polarization asymmetry, including a change of sign \cite{MAR00}. The difference between the two sets of predictions lies in the treatment of the non-resonant background terms: Janssen \textit{et al.} introduce hyperon resonances in the $u$-channel to counterbalance the strength produced by the Born terms in a physically relevant way. The calculations also differ in the choice for the hadronic form factor. 

For the $K^{+}\Lambda$ channel, the calculations in MAID2000 over-predict the beam polarization asymmetries and those by Janssen \textit{et al.} under-predict the measurements. For the $K^{+}\Sigma^{0}$ channel, the calculations predict similar absolute values for the beam polarization asymmetries, but with opposite sign. The measurements give positive values, but the magnitude is lower than the values by Janssen \textit{et al.} The discrepancy between the data and calculations does not necessarily mean that the models have fundamental shortcomings. It could merely indicate that the freedoms are too large and that fitting to cross section data only does not give sufficient boundary conditions. The photon polarization data presented here are great assets to guide the theoretical work.

For $E_{\gamma}>2.0$ GeV the above-mentioned models are no longer applicable. Regge-model calculations \cite{GUI97}, which reproduce the asymmetry at higher photon energies ($E_{\gamma}>5$ GeV) well, are not applicable for energies below $\sim2.5$ GeV since the $s$-channel resonances are not taken into account. The new data up to 2.4 GeV provide, therefore, another challenge for future theoretical work.

In short, we present beam polarization asymmetry data for the $p(\gamma,K^{+})\Lambda$ and $p(\gamma,K^{+})\Sigma^{0}$ reactions for $1.5<E_{\gamma}<2.4$ GeV and $0.6<\cos(\theta_{K^{+}}^{cm})<1.0$. Based on the calculations by Mart and Bennhold \cite{MAR00}, the positive sign measured in case of the former reaction indicates the presence of a missing $D_{13}$ resonance. However, in light of the large freedoms in the models, such strong conclusions are premature. Using the new results to constrain the calculations, similar to the case for $\pi$ photoproduction at lower energy, will lead to a strongly enhanced understanding of the reaction mechanisms and are pivotal for testing the presence of missing resonances.

\begin{acknowledgments}
The authors thank the staff at SPring-8 for providing excellent experimental conditions during the long course of the experiment. This research was supported in part by the Ministry of Education, Science, Sports and Culture of Japan, the National Science Council of the Republic of China (Taiwan) and the National Science Foundation.
\end{acknowledgments}

\bibliography{prlk}

\end{document}